\documentclass[12pt]{iopart}
\usepackage{iopams}
\usepackage{braket}
\usepackage{bm}
\usepackage{graphicx}
\usepackage{color}
\usepackage{textcomp}
\usepackage{soul}
\usepackage{comment}

\newcommand{\red}[1]{\textcolor{red}{#1}}

\begin{document}

\title{High-fidelity non-adiabatic cutting and stitching of a spin chain via local control}

\author{P. V. Pyshkin$^{1,2}$, E. Ya. Sherman$^{3,4}$, J. Q. You$^{5}$, Lian-Ao Wu$^{4,6}$}

\address{$^1$Institute for Solid State Physics and Optics, Wigner Research Centre,
	Hungarian Academy of Sciences, P.O. Box 49, H-1525 Budapest, Hungary}
\address{$^2$Beijing Computational  Science Research  Center,  Beijing 100084,  China}
\address{$^3$Department of Physical Chemistry, The University of the Basque Country UPV/EHU, 48080 Bilbao, Spain}
\address{$^4$IKERBASQUE, Basque Foundation for Science, 48011 Bilbao, Spain}
\address{$^5$Department of Physics, Zhejiang University, Hangzhou 310027,  China}
\address{$^6$Department of Theoretical Physics and History of Science, The University of the Basque Country UPV/EHU, 48080 Bilbao, Spain}

\ead{pavel.pyshkin@gmail.com}

\begin{abstract}

We propose and analyze, focusing on non-adiabatic effects, a technique 
of manipulating quantum spin systems based on local ``cutting'' and ``stitching''  of
the Heisenberg exchange coupling between the spins. 
This first operation is cutting of a bond separating a single spin from  
a linear chain, or of two neighboring bonds for a ring-shaped array of spins. We show that the 
disconnected spin can be in the ground state with a high fidelity even after a 
non-adiabatic process. Next, we consider inverse operation of stitching these bonds to increase the 
system size. 
We show that the optimal control algorithm can be found by using common numerical 
procedures with a simple two-parametric control function able to produce a
high-fidelity cutting and stitching. These results can be applied for manipulating ensembles of 
quantum dots, considered as prospective elements for quantum information technologies, and for design of 
machines based on quantum thermodynamics.  

\end{abstract}

\date{\today}
\submitto{\NJP}
\maketitle

\section{Introduction}

Theory of quantum control and its applications ~\cite{QuantControl,Quantum-Control-numerical-book} attract considerable attention 
of researchers. The abilities of producing and transforming quantum states on demand become 
important also in connection with research on quantum computation and information~\cite{Nielsen2000}. The problems in this 
field are usually related to the studies of the system evolution under time-dependent Hamiltonians 
and the resulting preparation of the desired final states with the maximum possible fidelity. 

In general, these problems can be formulated as follows. Assume that we impose an external 
control on a given quantum system in such a way that its initial Hamiltonian $H_i$ and the final 
Hamiltonian after time $T$, $H_f$, are known. We want 
to achieve the ground state of $H_f$ as a result of the unitary evolution governed by a time-dependent Hamiltonian 
$H(t)$, with $H(0) = H_i$ and $H(T) = H_f$, whose initial state is the ground state of~$H_i$. 
The adiabatic passage is a well-known way~\cite{Born1928} to realize this process. 
However, in order to satisfy the conditions of the adiabatic theorem, one has to implement the 
evolution of slow-varying Hamiltonian for a very long time, where decoherence may ruin the quantumness
of the system. 

One of the strategies  
for quantum control is the ``adiabaticity shortcut''~\cite{Shortcuts-Torrontegui-Muga-2013} 
to allow the design of the Hamiltonian $H(t)$ in such a way that the system terminates at the 
ground state of $H_{f}$ at a relatively short $T$. There are lots of proposals for making such 
adiabatic-like passages in a finite, short time; for instance, the proposal of 
transitionless quantum driving~\cite{Demirplak_adiab_drive, Berry-2009}, pulse/noise control~\cite{Jing2014, HQC-Pyshkin2016}, and 
invariant-based inverse engineering ~\cite{Shortcuts-Torrontegui-Muga-2013}. 
However, these proposals require the ability to control the {\em entire} quantum system~\cite{delCampo2012,Damski2014,Saberi2014}. For example, 
to cut a spin chain in a short time with ``adiabatic shortcut'', one has to control and change with 
time {\em each} single spin and {\em each} spin-spin interaction bond~\cite{Spin-cutting-PLA-Ren2017} 
in the chain. 

%Moreover, the adiabaticity implies that for each intermediate time  $0<t<T$, the state of the 
%evolving system is an instantaneous ground state of~$H(t)$, but this is not necessary for our goal. 

In the present paper, we investigate the shortcut via only {\em local control} of a complex quantum system
described by the Heisenberg model for interacting spins in an external magnetic field. 
First, we focus on the problem of ``cutting'' or disentangling of a such system into two parts.
Second, we consider inverse problem of stitching of two systems into one - the process which 
can be characterized as a quantum assembly. 
In general, the current adiabatic quantum computation 
or quantum annealing \cite{Das2008} is a multipartite stitching process, as implemented, for example, 
in $D$-wave quantum processors. 
The initial Hamiltonian $H_i$ may describe a non-interacting independent spin system, and with 
time evolution the system is stitched 
as a correlated entity described by the site-bond spin model $H_f$, such as the Ising model.
Our direct numerical simulations show that such 
kind of control can be achieved with a high fidelity in the non-adiabatic domain.
These quantum processes with a finite time duration can be also interesting as a non-adiabatic 
counterpart to a local quench dynamics of a suddenly cut and stitched spin chains~\cite{Stephan2011}, 
and for bond impurity chain cutting~\cite{Banchi2013}. Another related field is the  
quantum thermodynamics ~\cite{Apollaro2015}, or the physics of 
quantum heat machines~\cite{Alicki1979,Gelbwaser2014,Mukherjee2016,Azimi2014}, 
where one part of the spin chain can be considered as a working medium, connected to and 
disconnected from the other part, which plays the role of a thermal reservoir. 

In general, such a spin system can be realized in an ensemble of quantum dots, proposed as hardware elements for  
quantum information processing~\cite{Loss1998}  as well as in other spin-based systems with a similar prospective
\cite{Bose2003}. The single spin manipulation in a quantum dot has become a well-controllable procedure~\cite{Busl2013,Russ2017}
by now. The spin-spin interaction, strongly dependent on the electron tunneling rate between the dots 
can be controlled electrically by a fast gating of the tunneling 
channels ~\cite{Burkard1999}.

\section{Separation of a quantum system}

\subsection{General description: evolution and fidelities}

Here we describe the problem in general, without referring to a specific system. 
Let us assume that some joint quantum system consists of two parts: A and B. Initially, there is an interaction between A and B, 
such that the ground state~$\ket{\psi_0}$ of the joint A+B system generally cannot be presented 
as a direct product of A and B states: $\ket{\psi_0}\neq \ket{\varphi_{\rm {0A}}}\otimes\ket{\varphi_{\rm {0B}}}$, 
where $\ket{\varphi_{\rm {0A(B)}}}$ is the ground state of system A(B). Assume that initial Hamiltonian 
of this joint system has the form:
\begin{equation}\label{initial_H}
H_i = H_0 + V,
\end{equation}
where $V$ corresponds to the interaction between A and B parts. We can also write the 
following relation: $(H_0 + V)\ket{\psi_0} = \lambda_{\min} \ket{\psi_0}$, where $\lambda_{\min}$ 
is the minimal eigenvalue of the Hamiltonian~(\ref{initial_H}). Let us assume 
the ``switch-off'' of the $V$~interaction while the unitary evolution~$U(t)$ is governed by 
the time-dependent Hamiltonian $H(t) = H_0 + g(t)V$, with initial 
state $\ket{\psi(t=0)} = \ket{\psi_0}$, $g(t\leq0) = 1$, and $g(t\geq T) = 0$, 
where $T$ is the time of cutting. The state at time $t$ can be written as $\ket{\psi(t)} = U(t)\ket{\psi_0}$, where 
\begin{equation}\label{general_U}
U(t) = \mathcal{T}\exp\left[{-i\int_0^t H(s)ds}\right],
\end{equation}
with $\mathcal{T}$ denoting the time-ordering. Here and below we use the units with $\hbar\equiv 1$.
To characterize the evolution of the subsystems A and B we intoduce their reduced density matrices 
presented as traces over the other one: 
\begin{equation}
\rho_{\rm A(B)}(t) = \mathrm{tr_{\rm B(A)}}\ket{\psi(t)}\bra{\psi(t)}.
\label{rhoAB}
\end{equation}

\begin{figure}
	\centering
	\begin{minipage}{.5\textwidth}
		\centering
		\includegraphics*[width=.6\linewidth]{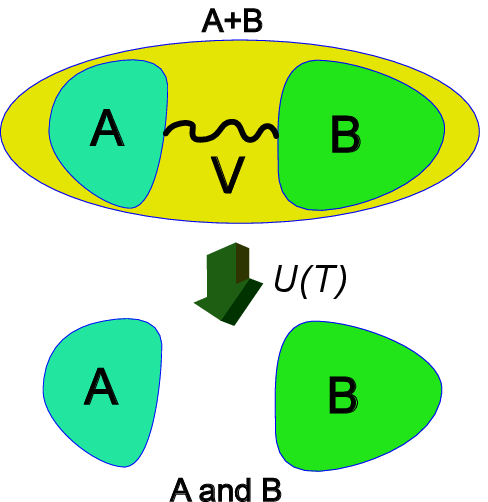}
		\caption{Schematic illustration of quantum system separation into two parts.}
		\label{fig1}
	\end{minipage}% 
	\hspace{-2cm}
	\begin{minipage}{.6\textwidth}
		\centering
		\includegraphics*[width=.4\linewidth]{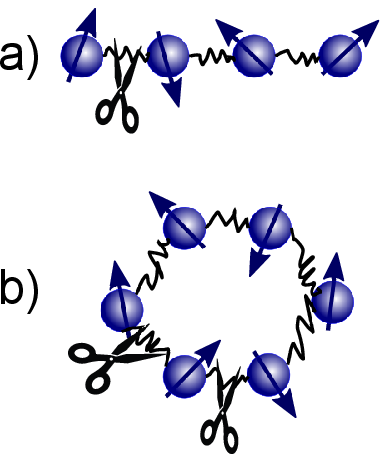}
		\caption{Schematic illustration of a spin system cutting: a) linear chain and b) ring-shaped array of spins.
		Subsystem A corresponds to the single spin. All other spins form subsystem B.}
		\label{fig2}
	\end{minipage}
\end{figure}

After the interaction~$V$ between A and B is switched off, we may consider A and B as 
two separate systems, which however, can be entangled. 
Our goal is to make the switch-off process in such a way that the system A is finally in its ground 
state (see figure \ref{fig1}) regardless of the final state of the system B. 
Although the sizes of these systems are different, their 
purities~$\mathrm{tr}\rho_{\rm A(B)}^{2}(t)$ and entropies 
$S_{\rm A(B)}(t)=-\mathrm{tr}\rho_{\rm A(B)}(t)\ln\rho_{\rm A(B)}(t)$
are equal to each other~\cite{Araki1970}. 

The final Hamiltonian can be written as $H(t\geq T) = H_0 = H_{\rm {0A}} + H_{\rm {0B}}$, 
where $H_{\rm {0A}}$ and $H_{\rm {0B}}$  have their own ground states 
$H_{\rm {0A(B)}}\ket{\varphi_{\rm {0A(B)}}} = \lambda_{\min\,\rm{A(B)}}\ket{\varphi_{\rm{0A(B)}}}$. 
To trace the proximity of A system to the ground state of $H_{\rm {0A}}$, 
we introduce time-dependent parameter 
\begin{equation}
f_{c}(t) = \sqrt{\braket{\varphi_{\rm {0A}} | \rho_{\rm A}(t) | \varphi_{\rm {0A}}}},
\label{fct}
\end{equation} 
and the resulting fidelity of the cutting process $f_{C}\equiv f_{c}(T).$ 

A complementary general parameter, describing the proximity of the state at given $t$ to the  
ground state of $H(t),$ $\ket{\psi_0(t)},$ is defined as:
\begin{equation}
{f}_{g}(t) =  |\braket{\psi_0(t)|\psi(t)}|, 
\label{ftilde}
\end{equation}
and corresponding fidelity ${f}_{G}\equiv{f}_{g}(T).$ 
It is easy to prove the inequality: $f_{C}\geq f_{G}$, such that the $f_{G}$ 
is the lower bound for $f_{C}$. 
We will use~$f_{G}$ to describe the reverse problem for controllably entangling 
initially separated A and B parts into the joint A+B system. 

Before proceeding to the numerical analysis,
we notice the importance of the commutator $[H_{0}, V].$  
It is easy to see from equation (\ref{general_U}) by using Magnus and Zassenhaus~\cite{magnus} 
formulas that if $[H_0, V] = 0$, the fidelities $f_{C}$  and ${f}_{G}$ do not 
depend on $g(t).$ Therefore, we assume that $[H_0, V] \neq 0$.

Since the propagator~(\ref{general_U}) for our problem is not presentable  
in a closed analytical form, one cannot produce an exact equation for $g(t)$ assuring a
shortcut to adiabaticity. In general, several numerical algorithms 
for optimal control of quantum systems have been proposed (see, e.g. \cite{krotov-book}) 
and implemented \cite{Krotov1-SOMLOI199385}. Recently, an optimized algorithm has been developed 
for the control pulses for practical qubits \cite{Machnes2018}.  

In this paper, we present numerical results of the successful optimization of the~$g(t)$ function by 
choosing appropriate set of parameters for its representation. The target fidelity~$f_{C}$ or $f_{G},$ being a 
functional of $g(t)$, becomes then a function of these parameters. We will show how this set 
can be optimized, study the properties of the optimal solutions, and provide a picture in basic physical and 
mathematical terms explaining main features of the proposed algorithm. 

\subsection{The model: Heisenberg chain in magnetic field}

We consider a chain with~$N$ spins (see figure \ref{fig2}) 
placed in a uniform external magnetic field with the following Hamiltonian
\begin{equation}\label{H_chain}
	H = J\sum_{n=1}^{N-1} \bm{\sigma}_n \bm\sigma_{n+1} + B \sum_{n=1}^N {\sigma}^{z}_n,
\end{equation}
where $\bm{\sigma}_{n}=({\sigma}_{n}^{x},{\sigma}_{n}^{y},{\sigma}_{n}^{z})$ is the Pauli matrix vector, $B$ is the 
external magnetic field along the $z$ direction, $J=\pm 1$ for (anti-) ferromagnetic coupling, and $n$ is the spin number. 
For the ring-shaped array we add $J{\bm\sigma}_{N}{\bm\sigma}_{1}-$term to equation ~(\ref{H_chain}) 
to assure the system periodicity. 
We consider the whole chain as the A+B system and the first spin as the A system (see figure \ref{fig2}). 
In such a case, we have 
$H_0 = J\sum_{n=2}^{N-1} \bm\sigma_n \bm\sigma_{n+1} + B \sum_{n=1}^N {\sigma}^{z}_n$ 
and $V = J\bm\sigma_1 \bm\sigma_{2},$ while for a ring-shaped array $V = J\bm\sigma_1 \bm\sigma_{2} + J\bm\sigma_1 \bm\sigma_{N}$.
After separating this single spin, we obtain $H_{\rm{0A}}=B{\sigma}^{z}_{1}$ with the eigenvalue $\lambda_{\min \rm{A}}=-B$
and spin-down eigenstate $\ket{\varphi_{\rm{0A}}}=\ket{\downarrow}_{1}$. 
Thus, we have satisfied $[H_0, V]\neq0,$ where the time dependence of the coupling is due to the time dependence 
of the exchange interaction in the corresponding bonds. 
It is obvious that for a ferromagnetic coupling with $J<0$, the 
ground state of the A+B system is disentangled \cite{Izyumov} 
with ~$\ket{\psi_0} = \ket{\varphi_{\rm {0A}}}\otimes\ket{\varphi_{\rm {0B}}}$ 
and therefore the cutting with the $f_{C}=1$ can be made instantaneously. Thus, we concentrate 
only on the nontrivial antiferromagnetic coupling $J>0$ and take $J=1$ as the energy unit.

 \begin{figure}
 	\begin{center}
 		\includegraphics{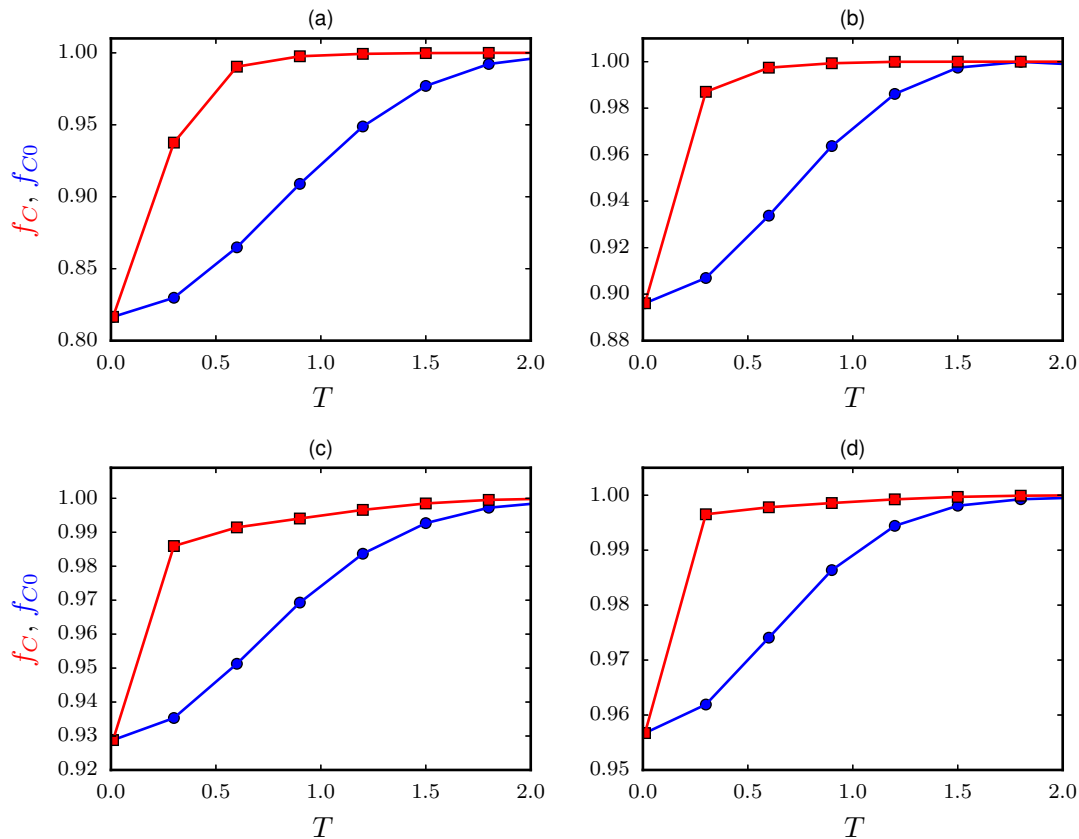}
 	\end{center}
 	\caption{Comparison of fidelities for optimized ($f_{C}$, red squares) and non-optimized 
 	($f_{C0}$, blue circles) cutting: ring-shaped array (a) $N=6$, (b) $N=7$; 
 	linear chain (c) $N=6$ and (d) $N=7$ with the difference 
 	between the results for odd- and even-$N$ being partially related to the 
 	formation of the spin singlets for even $N$ \cite{Izyumov}. 
 	Magnetic field $B=2$ and spin-spin interaction $J=1$.  }
 	\label{fig3}
 \end{figure}
 
\subsection{Parametrization of $g(t)$ and numerical results} 

We begin with a polynomial representation of $g(t)$ in the form:
\begin{equation} 
\label{g-represent}
g(t) = 
\cases{\displaystyle{1 + \sum_{n=1}^{K}a_n \left( \frac{t}{T} \right)^n},&  for $0\leq t \leq T,$ \\
0,& for $t > T,$ \\}
\end{equation}
where $a_{2},\dots,a_{K}$ are free parameters and $a_1 = -(1 + a_2 + a_3 + \dots + a_K)$. Thus, the fidelity 
is a function of $a_{2},\dots,a_{K}$ as $f_{C} \equiv f_{C}(a_2, a_3, \dots, a_K, T)$. In the realization 
with the simplest linear decrease of the $V$ term, where $a_{1}=-1,$ and $a_{n>1}=0,$ we define
\begin{equation}
\label{adiaresult}
f_{C0} \equiv f_{C}(0, 0, \dots, 0, T);\qquad \lim_{T\rightarrow\infty}f_{C}=\lim_{T\rightarrow\infty}f_{C0}=1.
\end{equation}
Our goal is to find a set of parameters $\mathbf{a} = \{a_2, a_3, \dots, a_K\}$ which maximizes the 
fidelity $f_C$ for a non-adiabatic process with a given finite time~$T$. To find the proper~$\mathbf{a}$, 
we use Broyden-Fletcher-Goldfarb-Shanno (BFGS) optimization method~\cite{BFGS}. 

\begin{figure}
 	\begin{center}
 		\includegraphics{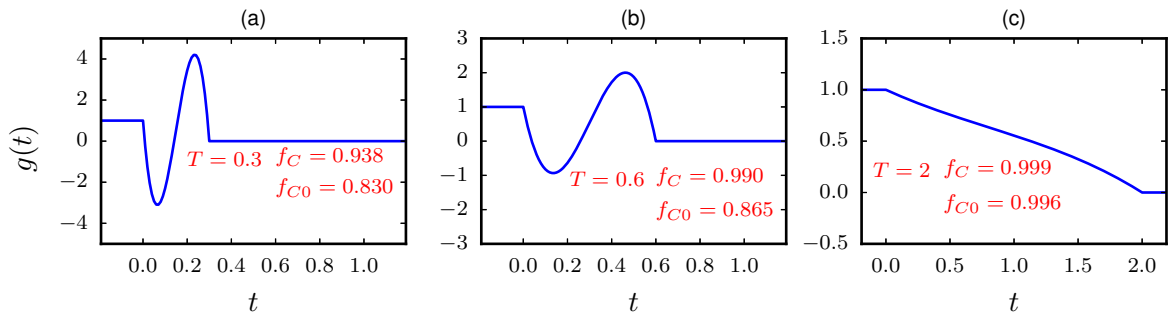}
 	\end{center}
 	\caption{Optimized control function~$g(t)$ for the ring-shaped array with $N=6$, $B=2$, $J=1$ 
 	for cutting times $T=0.3, 0.6, 2$. The shape of $g(t)$ corresponds to the parametrization 
 	given by~(\ref{g-represent}) with two free parameters. Fidelity~$f_{C0}$ corresponds to 
 	linear time dependence of ~$g(t)$.}
 	\label{fig-shape-param}
 \end{figure}

In figure \ref{fig3} we show $f_{C}$ for optimized cutting and $f_{C0}$ for non-optimized cutting as 
a function of $T$ for the ring-shaped and linear chains with~$N=6\, \mathrm{ and }\, 7$, $J=1$, and $B=2$. 
We use two free parameters $a_2$, and $a_3$, with~$K=3$ in equation ~(\ref{g-represent}). Numerical 
simulations have been performed by using a Python SciPy package with the built-in BFGS optimization method. 
Unitary operator in equation ~(\ref{general_U}) was approximated with $300$ time steps, 
and the error in fidelities is less than~$0.001$. 
Gradient for BFGS was approximated as a finite difference with a step of~$0.1$ in the parameter space. 
Note, that the magnetic field $B$ should not be very high, because in such a case the ground state becomes fully
spin polarized and disentangled. In figure \ref{fig-shape-param} we show the optimal shape of the control 
function~$g(t)$ for the ring-shaped array of spins. The linear chain has a similar shape of $g(t)$. 
Obviously, $g(t)$ becomes closer to the simple linear dependence with increasing the cutting time~$T$. 
Numerical data for this setting is presented in table~\ref{t1}. As can be seen from figure~\ref{fig3} and table~\ref{t1}, 
the same high value of $f_C$ can be achieved in two different ways: by a slow linear adiabatic 
turning off the interaction, and via a more complicated but faster switching. 
Thus this second way for achieving the demanded state can be considered as the adiabaticity shortcut.  

\begin{table}[]
	\centering
	\caption{Non-optimized and optimized fidelity for ring-shaped array of spins with $N=6$, $B=2$, $J=1$.}
	\begin{tabular}{l|l|l|l|l|}
		& $T=0.3$ & $T=0.6$ & $T=0.9$ & $T=2$ \\ \hline
		$f_{C0}$& 0.830 & 0.865 & 0.910 & 0.996 \\
		$f_{C}$ & 0.938 & 0.990 & 0.998 & 0.999\\
		$a_{2}$& 122.8 & 54.3 & 20.0 & 0.87 \\
		$a_{3}$& -82.0 & -36.3 & -13.5 & -0.72
	\end{tabular}
	\label{t1}
\end{table}

\begin{figure}
 	\begin{center}
 		\includegraphics{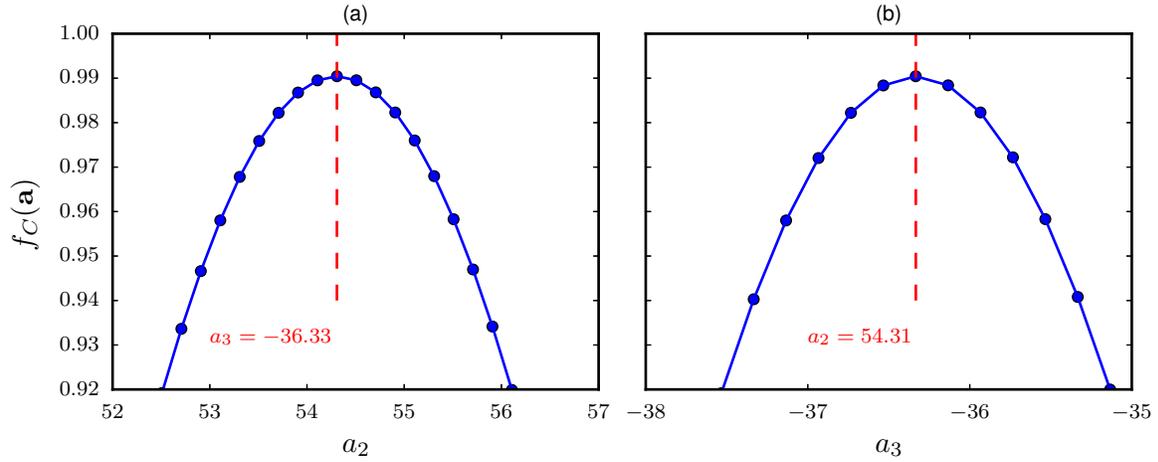}
 	\end{center}
 	\caption{Target fidelity as a function of variation of control parameters around its optimal 
 	values. Parameters are: ring-shaped array with $N=6$ spins, $B=2$, $J=1$, $T=0.6$ (second column in table~\ref{t1}) }
 	\label{fig5}
 \end{figure}

In figure \ref{fig5} we show $f_{C}$ as a function of the parameter $a_2$ ($a_3$) with a fixed  
at the optimal value $a_3$ ($a_2$). Here $T=0.6$ for the ring-shaped array with $N=6$, $J=1$, and $B=2$, where the
optimal values are $a_2 = 54.31$ and $a_3 = -36.33$ (the optimal values correspond to the red dashed 
vertical lines in figure \ref{fig5} and are shown in the second column in Table~\ref{t1}). 
As can be seen in figure \ref{fig5} when parameters $a_2$ and $a_3$ have a small deviation 
$(\approx2\%)$ from their optimal values, our proposal is still efficient. The parameter~$a_3$ must be 
tuned with more accuracy than parameter~$a_2$. 

\begin{figure}
 	\begin{center}
 		\includegraphics{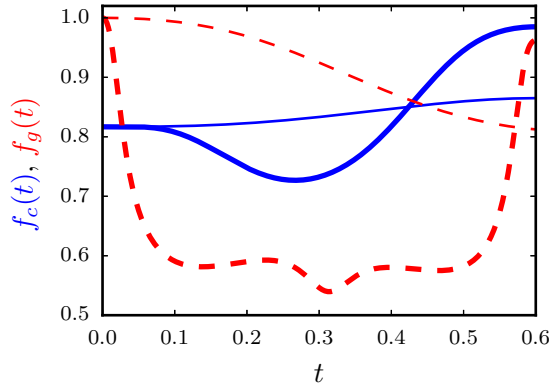}
 	\end{center}
 	\caption{Time dependence $f_{g}(t)$ (red dashed lines) and $f_{c}(t)$ (blue solid lines) for the 
 	optimized (bold) and non-optimized (thin) cutting. The system is a ring-shaped array of spins
 	with $N=6$, $B=2$, $J=1$ for cutting time $T=0.6$. }
 	\label{fig_non_adia}
 \end{figure}

In figure \ref{fig_non_adia}, we show the time-dependent 
$f_{g}(t)$ in equation (\ref{ftilde}) and $f_{c}(t)$ in equation (\ref{fct}) for optimized and 
non-optimized (linear) cutting protocols. One can see that while in the middle of the optimized 
process, $f_{c}(t)$ can be smaller than that for the non-optimized cutting, 
at $t=T$, it becomes higher. Moreover, as can be seen from the behavior of~$f_{g}(t)$, 
optimized process is less adiabatic than the non-optimized one during most of the evolution time. 
However, as mentioned above, we are interested only in the final state.

In addition, we note that even in the case of the energy-levels crossing in the cutting process (i.e. when at some time $t^{\prime}$
we have  $E_{0}(t^{\prime})=E_{1}(t^{\prime}),$
where $E_{0}(t)$ and $E_{1}(t)$  is the ground and the first excited state
energy of the instantaneous $H(t),$ respectively), which happens, 
e.g., in a ring-shaped array with $N=7$, $B=2$, $J=1$, the behavior of $f_{C}$ 
(or $f_{C0}$ corresponding to the linear switch-off) is the same as without the crossing. 
Fidelities $f_{C}$ and $f_{C0}$ increase up to $1$ with the evolution 
time~$T$ (see figure \ref{fig3}(b)). However, calculated $f_{G}=0$ here, meaning 
that A and B parts become disentangled after cutting, with A being in its ground state and  
B in a pure state orthogonal to the ground one. 

\subsection{Robustness of the control}

\begin{figure}
 	\begin{center}
 		\includegraphics{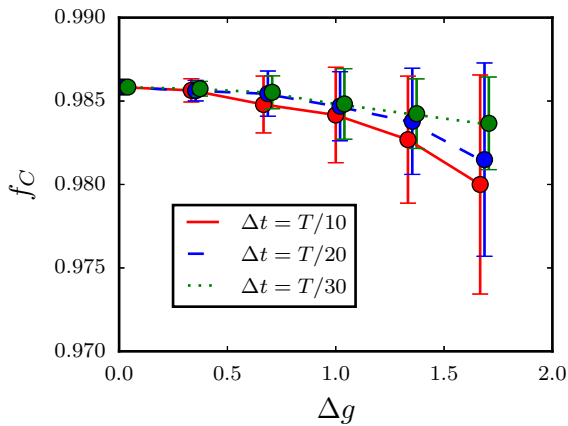}
 	\end{center}
 	\caption{The influence of noise with characteristic strength~$\Delta g$ to the fidelity~$f_C$ of 
 	cutting a linear spin chain. Error bars correspond to the standard deviation ($50$ random noise 
 	realization for each point was made). Here $N=6$, $B=2$, $J=1$, and $T=0.6$. }
 	\label{fig7}
 \end{figure}

Now we consider the influence of the ``apparatus'' noise \cite{Ban2012,Ulcakar2017} in the control function $g(t)$ on the 
fidelity of cutting. The noise is simulated as a set of rectangular pulses with a fixed length~$\Delta t$ 
and the random strength~$\Delta g (1/2 - r)$, where $\Delta g$ is a characteristic strength, and $r$ is a 
random number in the half-interval~$[0,1)$. This noise is added to the smooth $g(t)$ 
given by  equation ~(\ref{g-represent}) for the linear chain. 
The results of simulations presented in figure \ref{fig7} lead to conclusion that our proposal is 
robust against small fluctuations in the control function. Also, as can be seen in figure \ref{fig7}, a noise 
with a high characteristic frequency influences the fidelity less than that with a low frequency.

\subsection{Other $g(t)-$realizations}

\begin{figure}[h]
	\begin{center}
		\includegraphics{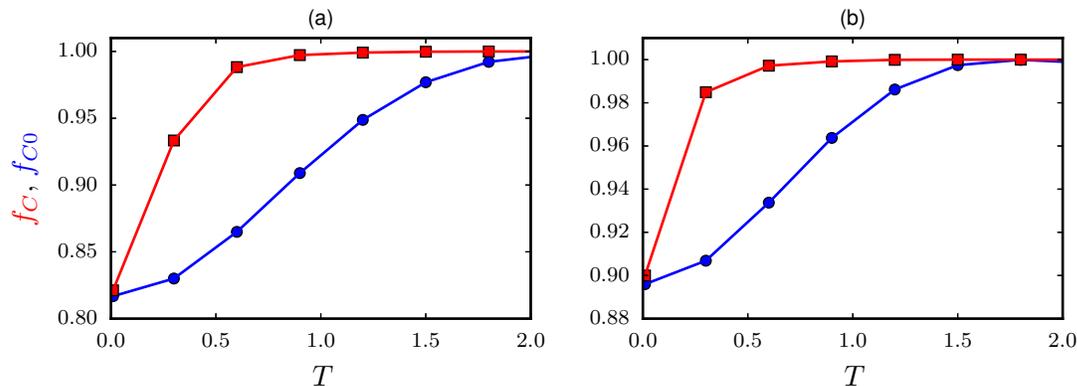}
	\end{center}
	\caption{Comparison of fidelities for optimized ($f_{C}$, red squares) and non-optimized ($f_{C0}$, blue circles) 
		cutting for ring-shaped array (a) $N=6$, (b) $N=7$. Magnetic field $B=2$ and spin-spin interaction $J=1$. 
		Equation~(\ref{g-represent-sin}) has been used for parametrization of $g(t).$}
	\label{fig_sin}
\end{figure}

Since we are not restricted in choosing the parametrization of the control function~$g(t)$, here we
compare the results of different representations. According to the 
boundary conditions, we can write~$g(t)$ as follows:

\begin{equation} \label{g-represent-sin}
g(t) = 
\cases{\displaystyle{1 -\frac{t}{T} + \sum_{n=1}^{K}b_n \sin\left(\frac{n \pi t}{T}\right)},&  for $0\leq t \leq T,$ \\
	0,& for $t > T,$ \\}
\end{equation}
where $b_{1},\dots, b_{K}$ are free parameters. To compare the parametrizations 
in equations (\ref{g-represent}) and (\ref{g-represent-sin}), 
we chose~$K=2$ in equation ~(\ref{g-represent-sin}) and repeat the optimization for the cutting of a ring-shaped array. 
The results are presented in figure \ref{fig_sin}. As can be seen in the figure,
the effectiveness of the optimization algorithm 
for different $g(t)$ representations is almost the same. 

\begin{figure}[h]
	\begin{center}
		\includegraphics{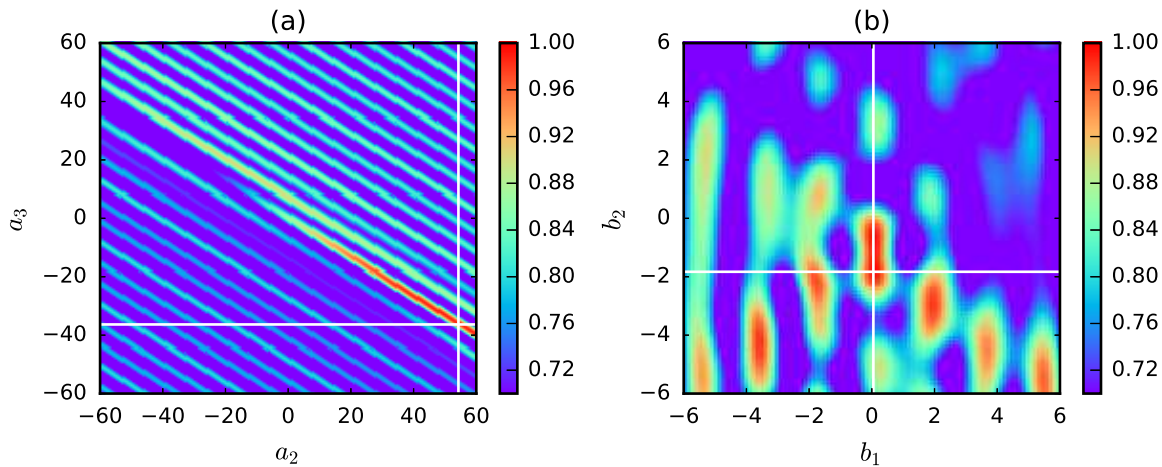}
	\end{center}
	\caption{Fidelity landscapes for cutting of ring-shaped array~($N=6$) as a function of two free parameters 
		for different types of parametrization: (a) equation (\ref{g-represent}) and (b) equation  (\ref{g-represent-sin}). 
		The time of cutting is~$T=0.6$, $B=2,$ and $J=1$. Vertical and horizontal white lines correspond to the 
		optimized parameters found by using BFGS optimization algorithm. The behavior of fidelity 
			at the cross-sections determined by the white lines in (a) is shown in figure (\ref{fig5}).
		%\blue{Red dotted lines in (b) corresponds to equations ${\frac{dg(t)}{dt}}|_{t=0} = 0$ and ${\frac{dg(t)}{dt}}|_{t=T} = 0$. 
		%Also, the shapes of control functions~$g(t)$ for parametrization~(\ref{g-represent}) and (\ref{g-represent-sin}) are quite similar.
		}
	\label{fig_lands}
\end{figure}

Since in our simulations we use the gradient-based BFGS optimization method, we do not need to calculate 
the fidelity~$f_{C}$ for each $\textbf{b}$ on a specially chosen dense grid. 
Nevertheless, it is instructive to look at the ``landscape'' of $f_{C}$ as a function of two free 
parameters and make sure that the optimization algorithm works correctly. Moreover, 
this data may be of independent interest. In figure \ref{fig_lands}, we show the landscapes  
corresponding to different parametrizations of~$g(t)$ and the same physical setting 
(ring-shaped array, $N=6$, $T=0.6$, $J=1$, and $B=2$). Intersections of white lines in figures  \ref{fig_lands}(a),(b)
are positioned at the optimal sets of parameters corresponding to the results of the BFGS algorithm. 
For both realizations, the optimal fidelities and control functions are very close to each other. 
It is interesting to observe that both the landscapes have a feature in common. The point corresponding to the 
best $f_{C}$ is located on the same high-fidelity ``island'' as the $(0,0)$-point, making it the right initial state 
for the global-minimum search algorithm, as it was employed in our simulations.

 \begin{figure} [h]
	\begin{center}
		\includegraphics{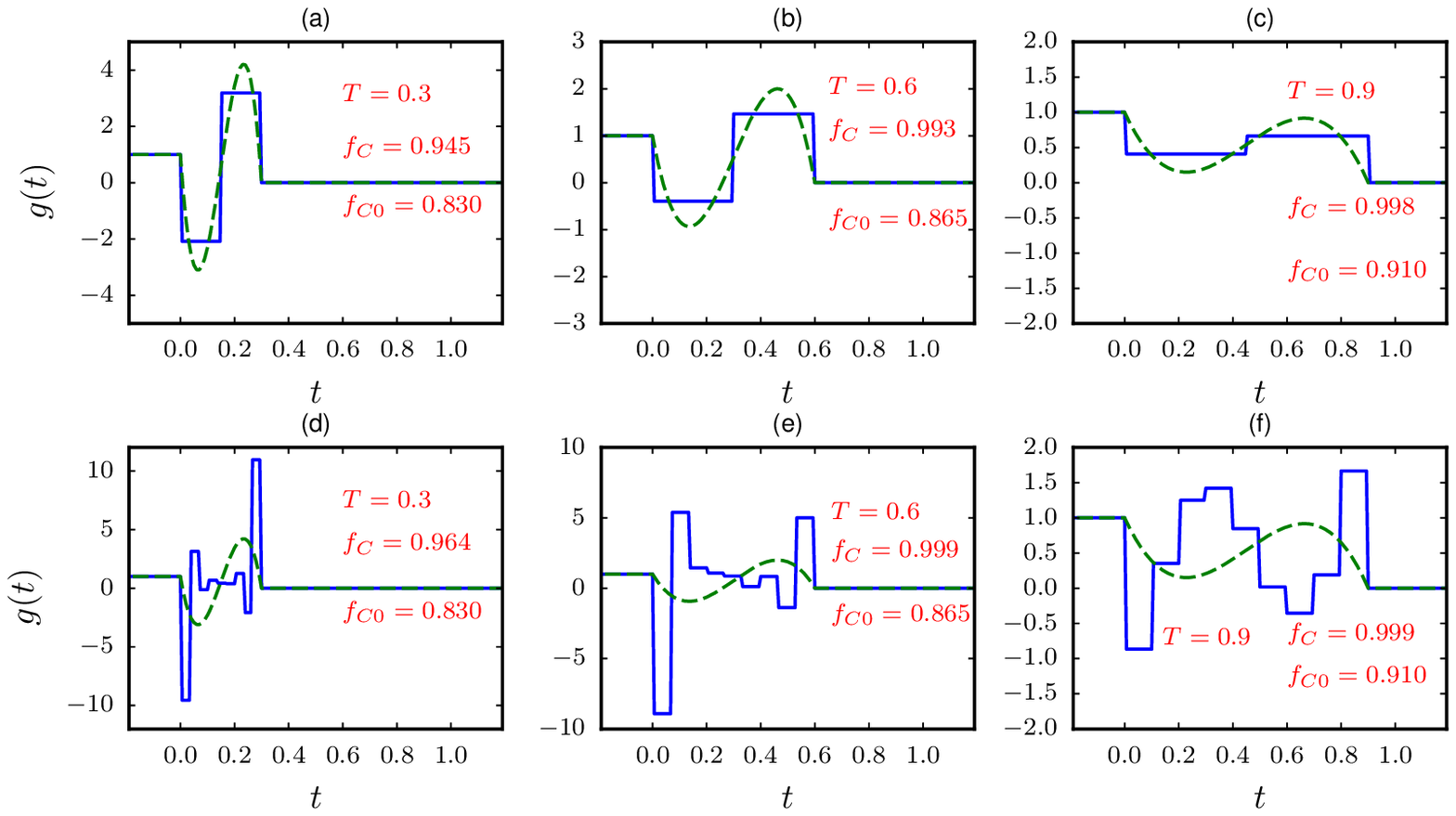}
	\end{center}
	\caption{Optimized control function~$g(t)$ for the ring-shaped array with $N=6$, $B=2$, $J=1$ for 
		cutting times $T=0.3, 0.6, 0.9$.  Subplots (a),(b),(c) corresponds to the pulse control~(\ref{g-pulsed}) 
		with two free parameters ($K=2$); subplots (d),(e),(f) corresponds to the pulse control~(\ref{g-pulsed}) with $K=9$. 
		Fidelity~$f_{C0}$ corresponds to the linear time dependence of $g(t)$. Dashed lines 
		show the optimal parametrization given by~(\ref{g-represent}) with two free parameters.}
	\label{fig4}
\end{figure}

We can also consider~$g(t)$ as a sequence of $K$ rectangular pulses whose amplitudes form the set of 
free parameters~$\mathbf{c} = \{c_1, \dots, c_K\}$, and the duration of each pulse is~$\Delta t = T/K$: 
\begin{equation}\label{g-pulsed}
g(t) = \sum_{n=1}^K c_n
\left[\theta \left(\vphantom{1^1}t - (n-1)\Delta t \right) - 
\theta \left(t - n\Delta t\right)     \right],
\end{equation}
where $\theta(t)$ is the Heaviside step function. 

In figures  \ref{fig4}(a) - \ref{fig4}(c), we show the 
optimal shape of $g(t)$ for~$K=2$, and in figures   \ref{fig4}(d)-\ref{fig4}(f) $K=9$ is used. 
Our setting is the same as in figure \ref{fig-shape-param}. By comparing the 
first row in figure \ref{fig4} with the shapes in figure \ref{fig-shape-param} (dashed lines), 
we see the correspondence between the shapes of smooth and pulse controls. Moreover, the optimization 
time for a pulsed shape with $K\approx10$ is much faster than the parametrized optimization with 
smooth shape of~$g(t)$ even for~$K=2$. This is because for optimization of a smooth $g(t),$ the algorithm divides the 
$[0,T]$ interval into more than $100$ pieces to accurately calculate the propagator~(\ref{general_U}) while for the pulse control 
the evolution interval is divided into the given number of $K$ pieces.

\section{Stitching of spin chain}

Now we consider the opposite process: a stitching of quantum system. In this setting, we have a 
disentangled initial state of the A+B system, which is the ground state of the Hamiltonian~$H_0$. 
Then, we switch on the interaction~$V$ between A and B systems, and the desirable final state of the joint A+B 
system is the ground state of the Hamiltonian~$H_0 + V$. This approach, 
which offers a new class of operations such as 
increasing the size of a quantum system, is of a practical importance, 
being a possible way of initializing a spin chain for 
quantum information processing (see e.g.~\cite{Schaller2008}).

%Moreover, a combination of cutting and stitching 
%can be used as a correction scheme of \blue{systematically} error-prone qubits by separating them  
%from the system and returning to it after the correction. 

Since both states of the A and B parts are important at the start and at the end of the stitching, we use 
the fidelity $f_{G}$ to describe this process. 
Thus we have to satisfy more strict condition in order to achieve the high fidelity: the energy 
gap between the two lowest states of the Hamiltonian $H = H_0 + g(t)V$ cannot be zero for any $g(t)\in (0,1)$. 
This requirement originating from the adiabatic theorem is not applicable to the chain 
cutting, where only the final state of the A system is essential. 

\begin{figure}
 	\begin{center}
 		\includegraphics{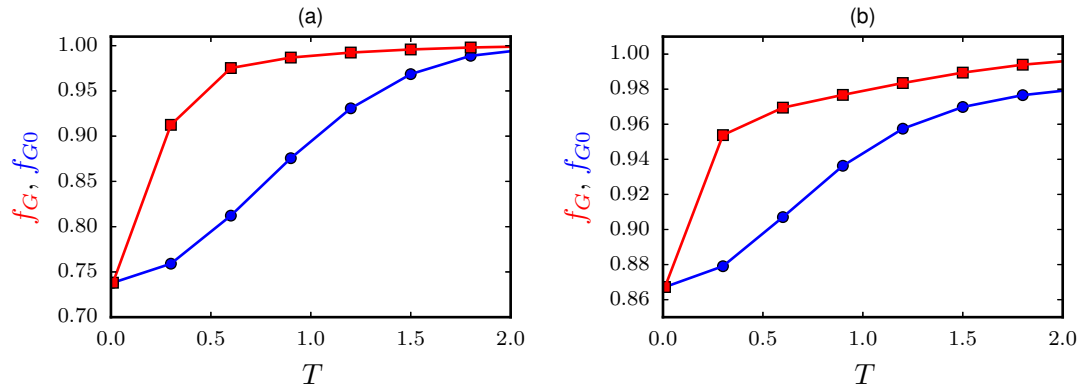}
 	\end{center}
 	\caption{Comparison of fidelities for optimized ($f_G$, red squares) and non-optimized ($f_{G0}'$, blue circles) 
 	stitching for ring-shaped array (a) $N=6$, (b) $N=7$. Magnetic field $B=2$ for (a) and $B=2.2$ for (b). 
 	Spin-spin interaction~$J=1$. }
 	\label{fig6}
 \end{figure}

Here we consider parametrization of the control function for the stitching process $g(t)$ 
expressed in the same way as for the cutting (\ref{g-represent}) by changing $t \rightarrow T - t$:
\begin{equation}\label{g-stich}
g(t) = 
\cases{\displaystyle{1 + \sum_{n=1}^{K}d_n \left( \frac{T - t}{T} \right)^n},&  for $0\leq t \leq T,$ \\
	1,& for $t > T$, \\}
\end{equation}
where $1 + \sum_{n=1}^{K}d_{n}=0.$ In addition, as in the case of cutting, 
we also use here the fidelity $f_{G0}$, similar to that defined in equation (\ref{adiaresult}). 
Note that the polynomial 
parametrization of $g(t)$ does not imply that the optimal control shape for stitching can be directly reconstructed 
from the optimal $g(t)$ for the corresponding cutting since the optimal sets of parameters
are essentially different for these processes. As an example we show  in figure \ref{fig6} the 
results for the optimization of the stitching process for a ring-shaped array.
 
Here we make a general comment related to the cutting and stitching processes.
The ground states of the initial and final Hamiltonians can be degenerate for the cutting or 
stitching process. In such a case we have the ground state subspace, and we have to choose the 
initial ground state $\ket{\psi_0}$ in such a way that $\braket{\psi_0|\psi_0'}\rightarrow1$ for $\ket{\psi_0'}$ 
being a non-degenerate ground state of the Hamiltonian $H = H_0 + (g(0) + \delta g)V$, where $g(0)=1$ 
and $\delta g\rightarrow -0$ ($g(0)=0$ and $\delta g\rightarrow +0$) for the cutting (stitching) process.

\section{Discussion}

\begin{figure}
	\begin{center}
		\includegraphics{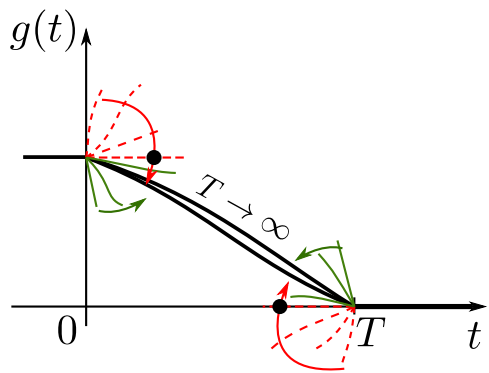}
	\end{center}
	\caption{ Schematic illustration of possible and impossible behavior of optimal 
	$g(t)$ near $t=0$ and $t=T$. Red dashed lines correspond to a 
	predetermined ineffective control and green solid lines correspond to a control, which 
	can be effective. Black solid lines belong to a family of controls, 
	which are all efficient in the limit~$T\rightarrow\infty$. Green and red arrows correspond 
	to changing the gradient~$dg/dt$ when the time $T$ goes from a finite value to  
	infinity. There intermediate points with $dg/dt=0$ (black dots) correspond to initially ineffective control. }
	\label{fig_gradientds}
\end{figure}

As can be seen from the shapes of control functions $g(t)$ in~figure \ref{fig4}, they have 
similar patterns for all cases of cutting, clearly seen for the fast processes. In all these patterns,
the interaction~$V$ enters a ferromagnetic domain with $g<0$ at the beginning, and jumps back to a 
strong anti-ferromagnetic coupling before the end of the process (see figure \ref{fig-shape-param} (a,b)). This behavior 
can be explained by the following heuristic arguments. In the adiabatic limit ($T\rightarrow\infty$) 
almost linear $g(t)-$dependence is the sufficient control for high output fidelity, as shown in figure \ref{fig-shape-param}(c). 
First, let us assume, that a non-adiabatic control which we have found for a finite $T=T_{0}$ has 
positive derivative $dg(+0)/dt>0$ in the case of a smooth control (see red dashed lines in figure \ref{fig_gradientds}), 
or the amplitude of first pulse~$c_{1}>1$ in the case of a pulse control (equation  (\ref{g-pulsed})). 
With the increase in $T,$ the shape of the control function must transform 
into a linear $g(t)$ continuously (red arrows in figure \ref{fig_gradientds}). 
This continuity implies that the derivative $dg(+0)/dt$ will become negative ~$dg(+0)/dt\rightarrow-1/T$ 
(or first pulse amplitude will go to $1 - 1/K$) with increasing time~$T$. Therefore, for some intermediate 
$T'>T_{0},$ the derivate should become zero, $dg(+0)/dt=0$ (or $c_{1}=1$). However, this vanishing derivate 
means that for this particular value $T=T'$ the control loses the efficiency, 
since at the beginning of the interval~$[0, T']$ it does not influence the system. 
On the contrary, if $dg(+0)/dt<0,$  for $T=T_0$ (or $c_{1}<1$) one {\em can} make continuous 
transformation of the control function to $T\rightarrow\infty$ without crossing such ``no-control'' 
point when ${dg(+0)}/{dt}=0$. 
The same reasoning can be made for the $g(t)-$behavior near the final $t=T$ point, 
where one must have~$dg(T-0)/dt<0$ or $c_{K}>0$
for the continuous and pulsed control, respectively. After analyzing all these options, 
we are left with the only possibility to have $dg(+0)/dt<0$ and $dg(T-0)/dt<0,$ as confirmed by our 
numerical simulations. This behavior of $g(t)$ for $T<1$ agrees well with the 
physical picture of formation of the 
spin ground state in the Zeeman field. The initial decrease in $g(t)$ to negative 
values forms a ferromagnetic coupling of the 
spin that is going to be separated (system A) with the nearest neighbor(s) from the 
rest of the chain (system B). This coupling 
decreases the quantum fluctuations in the corresponding Heisenberg spin-spin bond 
and the spins forming this bond together become easier oriented by the Zeeman field.

This general behavior of the control function is corroborated by the analysis of
figure~\ref{fig_lands} (b): the peak of the high-fidelity island is located at  a negative value of~$b_{2}$~parameter while $b_{1}$ is close to zero,
corresponding to a fast change in the sign of the exchange coupling at the corresponding bond. By using 
parametrization~(\ref{g-represent-sin}) and inequalities ${dg(+0)}/{dt}<0$ and ${dg(T-0)}/{dt}<0,$ one can analytically 
write the restrictions on $b_{1}$ and~$b_{2}$ as: (i)~$b_2 < 1/2\pi - b_1/2$ and (ii)~$b_2 < 1/2\pi + b_1/2$. 
Subsequent numerical analysis shows that the high-fidelity islands are located below the lines 
corresponding to these inequalities, in agreement with our reasoning. 
Note that the above reasoning is valid only if the requirements of the adiabatic 
theorem can be satisfied in the $T\rightarrow\infty$ limit, i.e. in the absence of the  
energy-levels crossings for $g\in[0,1]$.

The amplitude of $g(t)$ increases with decreasing of $T$, as expected from the 
time-energy uncertainty relation and quantum speed limit~\cite{MARGOLUS1998188}. However, it is easy 
to see that for effective control this amplitude should not be very large. 
To prove this, let us consider the pulse-shaped $g(t)$, where the evolution during the pulse 
can be described by~$U(t)\approx\exp[-i(g-1)Vt]$. Thus, due to the local nature of~$V$, the initial state 
cannot be changed to the demanded one by such evolution. Therefore, in the limit $T\rightarrow0$ ($\psi(T)\rightarrow\psi_{0}$),
the increase in $g(t)V$ cannot make the fidelity $f_{C}$  higher than the non-optimized~$f_{C0}.$ As a result,  
optimized and non-optimized curves merge in the anti-adiabatic limit $T\ll 1$ ($T=0.01$ in figure \ref{fig3}).

In general, our exact numerical simulations are limited by small number of spins $N$ 
since even for $N<10$ the optimization 
takes a relatively long time. However, the control function does not strongly depend on $N$ for 
long chains. Taking into account that the spin wave velocity (e.g. \cite{Cloizeaux1962,Ahmed2015}) is of the order of 
$J$, one can see that during the time $T$ the non-equilibrium 
spin wave propagate from the perturbed bond the distance (in the units of the spin-spin separation) 
of the order of $JT.$ Therefore, for 
$N\gg JT,$ the $g(t)-$ function should be only weakly $N-$dependent. However, the effect of parity, 
that is the difference between odd and even $N$ can influence the shape of $g(t).$

Although we presented here cutting protocols for detachment of a single spin from a chain, 
our results go far beyond preparation of a spin $1/2$ in the ground state. 
First, we demonstrated here the proof of concept of the high-fidelity 
non-adiabatic separation of a complex system, which is a new type of a quantum 
operation. Second, our proposal works well when we cut more than a single spin. 
For example, the non-optimized linear cutting of two spins from $N=5$ open 
chain with $B=2.1$ and $T=0.6$ yields $f_{C0}\approx0.26.$ However, our calculations 
demonstrate that under the pulse-shape control~(\ref{g-pulsed}) with $K=2$ and $c_{1}=-5.4$, $c_{2}=4.1$ 
the target fidelity becomes~$f_{C}\approx0.79$. Third, we note that 
cutting a single spin with the perfect fidelity $f_{C}\approx1$ means  
that the remaining part of the chain is in a known {\em pure} 
(not necessarily the ground) state, which in its turn can be transformed by unitary 
operations to a desirable pure state of $N-1$ spins. 
\begin{comment}
\red{The fidelity~$f(t)$ and $f'(t)$ are global characteristics of the cutting/stitching process, and now we propose a special kind of spin-spin correlation function~$E_{i,j}$ in order to analyze system dynamics under another point of view: 
\begin{equation}\label{E}
E_{i,j} = \braket{\bm{\sigma_i}\bm{\sigma_j}} - \braket{\bm{\sigma_i}}\braket{\bm{\sigma_j}},
\end{equation} 
where $i,j$ are the number of spins, and $\braket{\dots}$ is the quantum-mechanical average. Note, $E_{i,j}=0$ when spin $i$ is in A subsystem and spin $j$ is in B subsystem and the state of the whole A+B system is separable: $\ket{\psi} = \ket{\psi_{A}}\otimes\ket{\psi_{B}}$. Let us consider time dependencies $E_{12}(t)$, $E_{13}(t)$, $\dots$ for optimized and non-optimized processes where we cut the first spin.  }
\end{comment}

\section{Conclusion}

We have shown the feasibility of the unitary non-adiabatic cutting and stitching of complex quantum 
systems with a demanded high-fidelity output. The key feature of our proposal is the possibility of using {\em local} 
control instead of the global one.
The optimal shape of the control function can be found by the gradient numerical 
optimization. We show that even a simple two-parametric control can be effective for cutting and stitching the complex 
quantum system with a high fidelity. Since our approach is numerically exact, this proposal is directly applicable 
only for relatively small quantum systems. However, the increase in the system
size is not expected to modify our results qualitatively. Even when the time of the process is short, a
demanded fidelity can be achieved, and thus shortcut to adiabaticity can be realized. Different parametrizations of the control function can be used with a high fidelity of the results
robust against small variations of parameters and noise in the control. 
Our results can stimulate further investigations in the adiabaticity shortcut problems, including application in
quantum heat transfer and annealing processes.
From the experimental point of view, they can be realized by the electrical control of spin states in ensembles 
of quantum dots, where the modulation of spin-spin interaction, including change in the sign of the 
exchange coupling $J$ \cite{Burkard1999}
can be done by gating the interdot tunneling and the states inside the dots on the time scale 
as short as $10^{-2}$ nanoseconds, assuring a highly coherent evolution.   

\section*{ACKNOWLEDGMENTS}
This work was supported by the National Research, Development and Innovation Office of Hungary
(Project Nos. K124351 and 2017-1.2.1-NKP-2017-00001), Basque Government (Grant No. IT472-10), 
the Spanish Ministry of Economy, Industry, and Competitiveness (MINECO) 
and the European Regional Development Fund FEDER Grant No. FIS2015-67161-P (MINECO/FEDER, UE), 
and the NSFC No. 11774022.

\section*{References}

\bibliographystyle{iopart-num}

\bibliography{biblioteka}

\end{document}